\documentclass[12pt]{article}
\usepackage{epsf}

\catcode`@=12
\begin{document}
\thispagestyle{empty}
\begin{flushright} 
UCRHEP-T381\\ 
October 2004\
\end{flushright}
\vspace{0.5in}
\begin{center}
{\LARGE	\bf Grand Unification and Physics Beyond the Standard Model\\}
\vspace{1.5in}
{\bf Ernest Ma\\}
\vspace{0.2in}
{\sl Physics Department, University of California, Riverside, 
California 92521, USA\\}
\vspace{1.5in}
\end{center}

\begin{abstract}\
Recent progress in some selected areas of grand unification and 
physics beyond the standard model is reviewed.  Topics include gauge coupling 
unification, $SU(5)$, $SO(10)$, symmetry breaking mechanisms, finite field 
theory: $SU(3)^3$, leptonic color: $SU(3)^4$, chiral color and quark-lepton 
nonuniversality: $SU(3)^6$.
\end{abstract}
\vskip 1.0in

**Talk at V-SILAFAE, Lima, Peru (July 2004).

\newpage
\baselineskip 24pt

\section{Introduction}

Up to the energy scale of $10^2$ GeV, we are confident that the fundamental 
gauge symmetry of particle physics is that of the Standard Model (SM), i.e. 
$SU(3)_C \times SU(2)_L \times U(1)_Y$.  New physics may appear just above 
this scale, but there may also be a much higher energy scale where the 
three gauge groups of the SM become unified into some larger 
symmetry.  This is the notion of grand unification and depends crucially 
on the values of the three observed gauge couplings at the electroweak scale, 
as well as the particle content of the assumed theory from that scale to the 
unification scale.

\section{Gauge Coupling Unification}

The basic tool for exploring the possibility of grand unification is the 
renormalization-group evolution of the gauge couplings as a function of 
energy scale, given in one loop by
\begin{equation}
\alpha_i^{-1}(M_Z) = \alpha_i^{-1}(M_U) + (b_i/2\pi) \ln (M_U/M_Z),
\end{equation}
with the experimentally determined values $\alpha_3(M_Z) = 0.1183(26)$, 
$\sin^2 \theta_W(M_Z) = 0.23136(16)$, $\alpha^{-1}(M_Z) = 127.931(42)$, 
where $\alpha_2^{-1} = \alpha^{-1} \sin^2 \theta_W$, and $\alpha_1^{-1} = 
(3/5) \alpha^{-1} \cos^2 \theta_W$ (assuming $\sin^2 \theta_W (M_U) = 3/8$). 
The coefficients $b_i$ are obtained from the assumed particle content of 
the theory between $M_Z$ and $M_U$.  It is well-known that the three gauge 
coupings do not meet if only the particles of the SM are included.  However, 
if the SM is extended to include supersymmetry (MSSM) thereby increasing the 
particle content, they do meet at around $10^{16}$ GeV.

A recent detailed analysis\cite{bs04} using the more accurate two-loop 
analogs of Eq.~(1) shows that the MSSM does allow the unification of gauge 
couplings but there remains a possible discrepancy, depending on the choice 
of inputs at the electroweak scale.  In fact, this small discrepancy is 
taken seriously by proponents of specific models of grand unification, and 
has been the subject of debate in the past two years or so.

\section{$SU(5)$}

Consider the particle content of the MSSM.  There are three copies of 
quark and lepton superfields:
\begin{eqnarray}
&& (u,d) \sim (3,2,1/6), ~~ u^c \sim (3^*,1,-2/3), ~~ d^c \sim (3^*,1,1/3), \\ 
&& (\nu,e) \sim (1,2,-1/2), ~~ e^c \sim (1,1,1),
\end{eqnarray}
and one copy of the two Higgs superfields:
\begin{equation}
(\phi_1^0,\phi_1^-) \sim (1,2,-1/2), ~~ (\phi_2^+,\phi_2^0) \sim (1,2,1/2). 
\end{equation}
The quarks and leptons can be embedded into $SU(5)$ as follows:
\begin{eqnarray}
&& {\bf 5^*} = (3^*,1,1/3) + (1,2,-1/2), \\ 
&& {\bf 10} = (3,2,1/6) + (3^*,1,-2/3) + (1,1,1),
\end{eqnarray}
but the Higgs superfields do not form complete multiplets:  $\Phi_1 \subset 
{\bf 5^*}$, $\Phi_2 \subset {\bf 5}$.  Their missing partners are 
$(3^*,1,1/3)$, $(3,1,-1/3)$ respectively and they mediate proton decay.
In the MSSM, such effective operators are dimension-five, i.e. they are 
suppressed by only one power of $M_U$ in the denominator and can easily 
contribute to a proton decay lifetime below the experimental lower bound. 

Recalling that there is a small discrepancy in the unification of gauge 
couplings.  This can be fixed by threshold corrections due to heavy 
particles at $M_U$.  Using these heavy color triplet Higgs superfields to 
obtain exact unification, it was shown that\cite{mp02} their masses must lie 
in the range $3.5 \times 10^{14}$ to $3.6 \times 10^{15}$ GeV.  However, 
the experimental lower bound on the decay lifetime of $p \to K^+ \bar \nu$ 
is $6.7 \times 10^{32}$ years, which requires this mass to be greater than 
$7.6 \times 10^{16}$ GeV.  This contradiction is then used to rule out 
minimal $SU(5)$ as a candidate model of grand unification.

The above analysis assumes that the sparticle mass matrices are related to 
the particle mass matrices in a simple natural way.  However, proton decay 
in the MSSM through the above-mentioned dimension-five operators depends on 
how sparticles turn into particles.  It has been pointed out\cite{bfps02} 
that if the most general sparticle mass matrices are used, these operators 
may be sufficiently suppressed to avoid any contradiction with proton decay.  

Instead of adjusting the color triplet masses to obtain exact unification, 
a new and popular way is to invoke extra space dimensions.  For example, 
in a five-dimensional theory, if Higgs fields exist in the bulk, then 
there can be finite threshold corrections from summing over Kaluza-Klein 
modes.  A specific successful $SU(5)$ model\cite{hn02} was proposed using 
the Kawamura mechanism\cite{k00} of symmetry breaking by boundary conditions.

\section{$SO(10)$}

The power of $SO(10)$ is historically well-known.  A single spinor 
representation, i.e. {\bf 16}, contains the ${\bf 5^*}$ and {\bf 10} of 
$SU(5)$ as well as a singlet $N$, which may be identified as the right-handed 
neutrino.  The existence of three heavy singlets allows the three known 
neutrinos to acquire naturally small Majorana masses through the famous 
seesaw mechanism, and the decay of the lightest of them may also generate 
a lepton asymmetry in the early Universe which gets converted by sphalerons 
during the electroweak phase transition to the present observed baryon 
asymmetry of the Universe.

What is new in the past two years is the realization of the importance of 
the electroweak Higgs triplet contained in the {\bf 126} of $SO(10)$. 
Whereas the Higgs triplet under $SU(2)_R$ provides $N$ with a heavy 
Majorana mass, the Higgs triplet under $SU(2)_L$ provides $\nu$ with a 
small Majorana mass.\cite{ms98}  This latter mechanism is also seesaw 
in character and may in fact be the dominant contribution to the observed 
neutrino mass.  For a more complete discussion of this and other important 
recent developments in $SO(10)$, see the talk by Alejandra Melfo in these 
proceedings.
 
\section{Symmetry Breaking Mechanisms}

The breaking of a gauge symmetry through the nonzero vacuum expectation value 
of a scalar field is the canonical method to obtain a renormalizable field 
theory.  If fermions have interactions which allow them to pair up to form 
a condensate with $\langle \bar f f \rangle \neq 0$, then the symmetry is 
also broken, but now dynamically. With extra dimensions, a recent discovery 
is that it is possible in some cases for a theory without Higgs fields (in the 
bulk or on our brane) to be recast into one with dynamical symmetry breaking 
on our brane.  It is of course known for a long time that the components of 
gauge fields in extra dimensions may also be integrated over the nontrivial 
compactified manifold so that
\begin{equation}
\int A_i dx^i \neq 0,
\end{equation}
thereby breaking the gauge symmetry.\cite{h83}  More recently, bulk scalar 
field boundary conditions in a compact fifth dimension, using $S^1/Z_2 \times 
Z'_2$ for example,\cite{k00} have become the mechanism of choice for 
breaking $SU(5)$ and other grand unified groups to the MSSM.  This 
method can also be applied to breaking supersymmetry itself.\cite{ss79}

\section{Finite Field Theory: $SU(3)^3$}

If $\beta_i = 0$ and $\gamma_i = 0$ in an $N=1$ supersymmetric field theory, 
then it is also finite to all orders in perturbation theory if an isolated 
solution exists for the unique reduction of all couplings.\cite{lps88} This 
is an attractive possibility for a grand unified theory between the 
unification scale and the Planck scale.  The conditions for finiteness are 
then boundary conditions on all the couplings of the theory at the 
unification scale where the symmetry is broken, and the renormalization-group 
running of these couplings down to the electroweak scale will make 
predictions which can be compared to experimental data. In particular, the 
mass of the top quark and that of the Higgs boson may be derived.  Successful 
examples using $SU(5)$ already exist.\cite{kmz93,kkmz98}  Recently, an 
$SU(3)^3$ example has also been obtained.\cite{mmz04}

Consider the product group $SU(N)_1 \times ... \times SU(N)_k$ with $n_f$ 
copies of matter superfields $(N,N^*,1,...,1) + ... + (N^*,1,1,...,N)$ in a 
``moose'' chain.  Assume $Z_k$ cyclic symmetry on this chain, then
\begin{equation}
b = \left( -\frac{11}{3} + \frac{2}{3} \right) N + n_f \left( \frac{2}{3} + 
\frac{1}{3} \right) \left( \frac{1}{2} \right) N = -3N + n_f N.
\end{equation}
Therefore, $b=0$ if $n_f=3$ independent of $N$ and $k$.

Choose $N=3$, $k=3$, then we have the trinification model,\cite{dgg84} 
i.e. $SU(3)^3$ which is the maximal subgroup of $E_6$.  The quarks and 
leptons are given by $q \sim (3,3^*,1)$, $q^c \sim (3^*,1,3)$, and $\lambda 
\sim (1,3,3^*)$, denoted in matrix notation respectively as
\begin{eqnarray}
\pmatrix{d & u & h \cr d & u & h \cr d & u & h}, ~~~  
\pmatrix{d^c & u^c & h^c \cr d^c & u^c & h^c \cr d^c & u^c & 
h^c}, ~~~ 
\pmatrix{N & E^c & \nu \cr E & N^c & e \cr \nu^c & e^c & S}. 
\end{eqnarray}
With three families, there are 11 invariant $f$ couplings of the form 
$\lambda q^c q$ and 10 invariant $f'$ couplings of the form $\det q + 
\det q^c + \det \lambda$.  An isolated solution of $\gamma_i=0$ is
\begin{equation}
f^2_{iii} = \frac{16}{9} g^2,
\end{equation}
and all other couplings = 0.  Assuming that $SU(3)^3$ breaks down to 
the MSSM at $M_U$, this predicts $m_t \sim 183$ GeV, in good agreement with 
the present experimental value of $178.0 \pm 2.7 \pm 3.3$ GeV.

\section{Leptonic Color: $SU(3)^4$}

Because of the empirical evidence of gauge coupling unification, almost 
all models of grand unification have the same low-energy particle content 
of the MSSM, including all models discussed so far.  However, this does not 
rule out the possibility of new physics (beyond the MSSM) at the TeV energy 
scale, without spoiling unification.  I discuss two recent examples.  The 
first\cite{bmw04} is nonsupersymmetric $SU(3)^4$ and the second\cite{ma04} 
is supersymmetric $SU(3)^6$.

In trinification, quarks and leptons are assigned asymmetrically in Eq.~(9). 
To restore complete quark-lepton interchangeability at high energy, an 
$SU(3)^4$ model of quartification\cite{bmw04} has been proposed.  The idea 
is to add leptonic color\cite{fl90} $SU(3)_l$ which breaks down to 
$SU(2)_l \times U(1)_{Y_l}$, with the charge operator given by
\begin{equation}
Q = I_{3L} + I_{3R} - \frac{1}{2} Y_L - \frac{1}{2} Y_R - \frac{1}{2} Y_l.
\end{equation}
The leptons are now $(3,3^*)$ under $SU(3)_L \times SU(3)_l$ and $(3,3^*)$ 
under $SU(3)_l \times SU(3)_R$, i.e.
\begin{equation}
l \sim \pmatrix{x_1 & x_2 & \nu \cr y_1 & y_2 & e \cr z_1 & z_2 & 
N}, ~~~ l^c \sim \pmatrix{x^c_1 & y^c_1 & z^c_1 \cr 
x^c_2 & y^c_2 & z^c_2 \cr \nu^c & e^c & N^c}.
\end{equation}
The exotic particles $x,y,z$ and $x^c,y^c,z^c$ have half-integral charges:
$Q_x = Q_z = Q_{y^c} = 1/2$ and $Q_{x^c} = Q_{z^c} = Q_y = -1/2$, hence 
they are called ``hemions''.  They are confined by the $SU(2)_l$ ``stickons'' 
to form integrally charged partciles, just as the fractionally charged 
quarks are confined by the $SU(3)_q$ gluons to form integrally charged 
hadrons.

The particle content of $SU(3)^4$ immediately tells us that if unification 
occurs, then $\sin^2 \theta_W (M_U) = \sum I_{3L}^2/\sum Q^2 = 1/3$ instead 
of the canonical 3/8 in $SU(5)$, $SU(3)^3$, etc.  This means that it cannot 
be that of the MSSM at low energy.  Instead the SM is extended to include 
3 copies of hemions at the TeV scale:
\begin{equation}
(x,y) \sim (1,2,0,2), ~~ x^c \sim (1,1,-1/2,2), ~~ y^c \sim (1,1,1/2,2),
\end{equation}
under $SU(3)_C \times SU(2)_L \times U(1)_Y \times SU(2)_l$, without 
supersymmetry.  In that case, it was shown\cite{bmw04} that the gauge 
couplings do meet, but at a much lower unification scale $M_U \sim 4 \times 
10^{11}$ GeV.  However, proton decay is suppressed by effective 
higher-dimensional Yukawa couplings with $\tau_p \sim 10^{35}$ years. 
Also, the exotic hemions at the TeV scale have $SU(2)_L \times U(1)_Y$ 
invariant masses such as $x_1 y_2 - y_1 x_2$, so that their contributions to 
the $S,T,U$ oblique parameters are suppressed and do not spoil the agreement 
of the SM with precision electroweak measurements.

\section{Chiral Color and Quark-Lepton Nonuniversality:\\ $SU(3)^6$}

Each of the $SU(3)$ factors in supersymmetric unification may be extended:
\begin{equation}
SU(3)_C \to SU(3)_{CL} \times SU(3)_{CR},
\end{equation}
which is the notion of chiral color;\cite{fg87}
\begin{equation}
SU(3)_L \to SU(3)_{qL} \times SU(3)_{lL},
\end{equation}
which is the notion of quark-lepton nonuniversality;\cite{gjs89,lm03} and
\begin{equation}
SU(3)_R \to SU(3)_{qR} \times SU(3)_{lR},
\end{equation}
which is needed to preserve left-right symmetry.  Quarks and leptons are 
now $(3,3^*)$ under $SU(3)_{CL} \times SU(3)_{qL}$, $SU(3)_{qR} \times 
SU(3)_{CR}$, and $SU(3)_{lL} \times SU(3)_{lR}$.  The three extra 
$(3,3^*)$ multiplets $x,x^c,\eta$ transform under $SU(3)_{qL} \times 
SU(3)_{lL}$, $SU(3)_{lR} \times SU(3)_{qR}$, $SU(3)_{CR} \times SU(3)_{CL}$ 
respectively, with $x,x^c$ having the same charges as $\lambda$ and zero 
charge for $\eta$.  With this assignment, $\sin^2 \theta_W (M_U) = 3/8$.

Because all the fermions are arranged in a moose chain, this model is 
automatically free of anomalies, in contrast to the case of chiral color 
by itself or quark-lepton nonuniversality by itself, where anomalies 
exist and have to be canceled in some {\it ad hoc} way.  At the TeV scale, 
the gauge group is assumed to be $SU(3)_{CL} \times SU(3)_{CR} \times 
SU(2)_{qL} \times SU(2)_{lL} \times U(1)_Y$ with the following 3 copies 
of new supermultiplets:
\begin{eqnarray}
&& h \sim (3,1,1,1,-1/3), ~~ h^c \sim (1,3^*,1,1,1/3), ~~ \eta \sim 
(3^*,3,1,1,0); \\ 
&& (\nu_x,e_x) \sim (1,1,2,1,-1/2), ~~ (e^c_x,\nu^c_x) \sim (1,1,1,2,1/2); \\ 
&& \pmatrix{N_x & E_x^c \cr E_x & N_x^c} \sim (1,1,2,2,0).
\end{eqnarray}
Again they all have $SU(2)_L \times U(1)_Y$ invariant masses.  With this 
particle content, it was shown\cite{ma04} that unification indeed occurs 
at around $10^{16}$ GeV.  What sets this model apart from the MSSM is 
the rich new physics populating the TeV landscape.  In addition to the 
particles and sparticles listed above, the heavy gauge bosons and fermions 
corresponding to the breaking of chiral color to $QCD$ as well as 
quark-lepton nonuniversality to the usual $SU(2)_L$ should also be manifest, 
with unmistakable experimental signatures.

The consequences of $SU(2)_{qL} \times SU(2)_{lL} \to SU(2)_L$ have been 
discussed\cite{lm03} in some detail.  They include the prediction 
$(G_F)_{lq} < (G_F)_{ll}$, which may be interpreted as the apparent 
violation of unitarity in the quark mixing matrix, i.e.
$|V_{ud}|^2 + |V_{us}|^2 + |V_{ub}|^2 < 1$,
as well as effective $\sin^2 \theta_W$ corrections in processes such as 
$\nu q \to \nu q$, polarized $e^- e^- \to e^- e^-$, and the weak charge 
of the proton, etc.  However, the constraints from $Z^0$ data imply 
that these effects are very small and not likely to be measurable within 
the context of this model.  On the hand, since the new particles of this 
model are required to be present at the TeV scale, they should be 
observable at the Large Hadron Collider (LHC) when it becomes operational 
in a few years.

\section{Conclusion}

Assuming a grand desert from just above the electroweak scale to $10^{16}$ 
GeV, the particle content of the MSSM allows the unification of the three 
known gauge couplings. If studied closely, taking into account proton decay 
and neutrino masses, etc., this appears to favor $SO(10)$ as the grand 
unified symmetry over $SU(5)$ but the latter is still viable, especially 
if a fifth dimension is invoked for example.

Instead of a single simple group, the product $SU(N)^k$ supplemented by 
a cyclic $Z_k$ discrete symmetry is an interesting alternative.  Using 
a moose chain in assigning the particle content of such a supersymmetric 
theory, a necessary condition for it to be finite is to have 3 copies of 
this chain, i.e. 3 families of quarks and leptons.  A realistic example 
has been obtained\cite{mmz04} using $N=k=3$.

For $N=3$, $k=4$ without supersymmetry, the notion of leptonic color 
which has a residual unbroken $SU(2)_l$ gauge group can be implemented 
in a model of $SU(3)^4$ quartification\cite{bmw04}.  This model allows 
unification at $10^{11}$ GeV without conflicting with proton decay, and 
predicts new half-integrally charged particles (hemions) at the TeV scale. 

For $N=3$, $k=6$ with supersymmetry, the notions of chiral color and 
quark-lepton nonuniversality can be implemented\cite{ma04}, which cooperate 
to make the theory anomaly-free and be observable at the TeV scale, 
without spoiling unification.

In a few years, data from the LHC will tell us if the MSSM is corrrect 
[as predicted for example by $SU(5)$ and $SO(10)$], or perhaps that 
supersymmetry is not present but other new particles exist [as predicted 
for example by $SU(3)^4$], or that there are particles beyond those of 
the MSSM as well [as predicted for example by $SU(3)^6$].  Excitement awaits.

\section*{Acknowledgments}

I thank Javier Solano and all the other organizers of V-SILAFAE for their 
great hospitality and a stimulating meeting in Peru.  This work was supported 
in part by the U.~S.~Department of Energy under Grant No.~DE-FG03-94ER40837.

\end{document}